\documentclass[pra,twocolumn,  noshowkeys,  preprintnumbers,amsmath,amssymb, showpacs, aps, 10pt]{revtex4-1}
\usepackage{graphicx,color}
\usepackage{subfigure}

\usepackage{amsmath,enumerate}
\usepackage{bbm,amsfonts}

\newcommand{\tmop}[1]{\ensuremath{\operatorname{#1}}}

\newcommand{\ket}[1]{\left|#1\right\rangle}
\newcommand{\bra}[1]{\left\langle#1\right|}

\newcommand{\hideit}[1]{}

\begin{document}

\title{Quantum memories based on engineered dissipation}

\author{Fernando \surname{Pastawski}}
\author{Lucas \surname{Clemente}}
\author{Juan Ignacio \surname{Cirac}}
\affiliation{Max-Planck-Institut f{\"{u}}r Quantenoptik,
Hans-Kopfermann-Str.\ 1, D-85748 Garching, Germany.}

\begin{abstract}
Storing quantum information for long times without disruptions is a major requirement for most quantum information technologies.
A very appealing approach is to use {\it self-correcting} Hamiltonians, i.e. tailoring local interactions among the qubits such that when the system is weakly coupled to a cold bath the thermalization process takes a long time.
Here we propose an alternative but more powerful approach in which the coupling to a bath is engineered, so that dissipation protects the encoded qubit
against more general kinds of errors. We show that the method can be implemented locally in four dimensional lattice geometries by means of a toric code, and propose a simple 2D set-up for proof of principle experiments.
\end{abstract}

\maketitle

\section{Introduction}
There are two existing approaches to providing coherent quantum storage on
many-body systems. The first one corresponds to \emph{fault tolerant quantum circuits}\cite{shor_scheme_1995, gottesman_theory_1998}. 
If one can perform quantum gates and provide fresh initialized qubits with a sufficiently high accuracy and frequency, then quantum computing and in particular, quantum memory is possible for a time exponential in the dedicated resources.

More recently, Kitaev \cite{dennis_topological_2002, kitaev_fault-tolerant_2003} proposed that it might be possible to protect quantum information passively by engineering of suitable Hamiltonian systems, in analogy to magnetic domains for classical memories.
While an energetically degenerate code subspace insensitive to Hamiltonian perturbations is a necessary condition, it has become clear that there are additional requirements for this approach to quantum memories to work.
Possibly the most important requirement is to cope with the undesired coupling between the storage system and its environment.
In this direction, the approach that has benefited from the most theoretical progress  goes by the moniker of {\it self--correcting}  Hamiltonians, \cite{bacon_operator_2006, alicki_thermal_2008, hamma_topological_2008,  chesi_self-correcting_2009}.

For {\it self--correcting} Hamiltonians, a weak local coupling to a thermal bath is assumed. Making a Born-Markov approximation, the evolution of the system can be described by a thermalizing master equation. While for general local couplings, any initial state will decay to the unique Gibbs state, it is still possible for the decay rate of specific observables to become smaller as the number $N$ of subsystems increases. 
This leads to the possibility of storing quantum information by encoding it on a pair of slowly decaying anticommuting many-body observables. A Hamiltonian will thus be called {\it self--correcting} provided that below a certain finite bath temperature the dissipative dynamics leads to information lifetimes growing with the system size (typically following an exponential increase).
Alicky et al. \cite{alicki_thermal_2008} rigorously proved an exponentially long relaxation time for protected observables in the 4D toric code. 
Chesi et al. \cite{chesi_thermodynamic_2010} generalized this result deriving a criteria for quantum memory based on {\it self--correcting} Hamiltonians and lower bounds on the storage times.
However, it is in general not known how non thermal noise or even thermalization under a perturbed Hamiltonian \cite{pastawski_limitations_2010} affects this lifetime. In particular, this may be the case whenever the qubits are weakly coupled to an additional bath which induces a small rate of depolarization\cite{pastawski_long_2009}.

Building on previous results, we propose and analyze an alternative way of protecting quantum states. 
The method is similar to that of protecting Hamiltonians, but now the main idea is to tailor the coupling of the qubits to a bath, so that the engineered dissipation extends the life-time of the encoded qubit.
Apart from being passive (i.e. not requiring the burden of interrogating the quantum memory at intermediate times), the main advantage of this scheme is that it can potentially correct for other kinds of errors beyond those generated by thermalization, including depolarizing noise. 
In particular, we propose a specific method in 4 spatial dimensions inspired by toric codes and obtain evidence of its performance with the help of numerical simulations. 
We also investigate a simplified 2-dimensional model protecting only from phase errors which could be a good candidate for proof of principle experiments. 

Many-body classical memories based on dissipation (often under the name of {\it asynchronous celular automata}) have naturally appeared in the context of {\em classical} fault tolerant computation. For example, using a simple local update rule on a 2D lattice, Toom \cite{Toom1974, gray_Tooms_1999} showed that classical information can be protected against weak local noise. 
A more elaborate update rule by G\'acs \cite{gacs_reliable_2001} provide protection even on a 1D lattice. 
These results already suggest that dissipation may offer a powerful alternative to the existing methods for constructing many-body quantum memories, as investigated in the present work. 
In fact, several authors have already proposed the use of continuous quantum error correcting codes
\cite{paz_continuous_1998,ahn_continuous_2002, sarovar_continuous_2005, oreshkov_continuous_2007,mabuchi_continuous_2009}. 
However previous works concentrate on a single level of error correction and do not address the large $N$ many-body scenario. 
A notable exception is the work of Dennis et al. \cite{dennis_topological_2002} introducing a {\it heat bath algorithm} (thermal dissipation for the 4D toric code) in order to simplify the efficacy analysis of a local many body quantum error correction algorithm.
At the crux of this approach is that thermal dissipation can be interpreted not only as introducing decoherence (errors), but also as performing a form of error correction, with the balance between the two effects roughly given by the bath temperature. 
Indeed, this heat bath algorithm can already be seen as a 
dissipative quantum memory lending itself to more natural engineering. 
In fact, engineered dissipation is more general in that it need not satisfy detailed balance conditions and thus its power extends that of cooling a {\it self--correcting} Hamiltonian. 
In other words, the steady state need not be an equilibrium state and its dynamics may show a net flow (imagine a funnel receiving water from a hose). 
As the classical results show, this more general kind of dissipation may be crucial in order to correct general kind of errors.

Our proposal can be viewed as another example where engineered dissipation may become a useful and alternative tool in the context of quantum information processing, beyond quantum computation \cite{verstraete_quantum_2009}, state engineering \cite{verstraete_quantum_2009,diehl_quantum_2008}, or entanglement creation \cite{krauter_entanglement_2010}. 
In all those cases, it is desirable to be able to couple small subsets of qubits to Markovian environments so that their evolution equation follows a prescribed master equation.
As exposed in \cite{verstraete_quantum_2009}, {\em dissipative gadgets} provide a direct way of implementing this is in terms of {\em damped qubits}; that is, a set of qubits which themselves follow a damping master equation due to their coupling to an environment. 
Those qubits can be directly coupled to the physical qubits of the quantum memory or computer to provide the desired dissipation, and thus appear as an important resource in dissipative quantum information processing. 

This paper is organized as follows. In Section II we briefly present the general idea of engineered dissipative quantum memories. In Section III we display two different but rather obvious approaches to dissipative quantum memories and discuss why they are not entirely satisfactory. In Section IV we present a specific method in 4 spatial dimensions as well as the results of numerical simulations which validate the performance of the scheme. Section V contains a simplified version in 2 spatial dimensions which corrects against phase errors and that could be tested experimentally in the near future. In Section VI we show how one can use dissipative qubits to engineer the dissipation and analyze under which condition one can use them in this context. All previous section contain the main statements of our work. 
The detailed proofs of our results and more thorough explanations are given in the appendices.

\begin{figure}[h]
\begin{center}
\includegraphics[width=1.0\columnwidth]{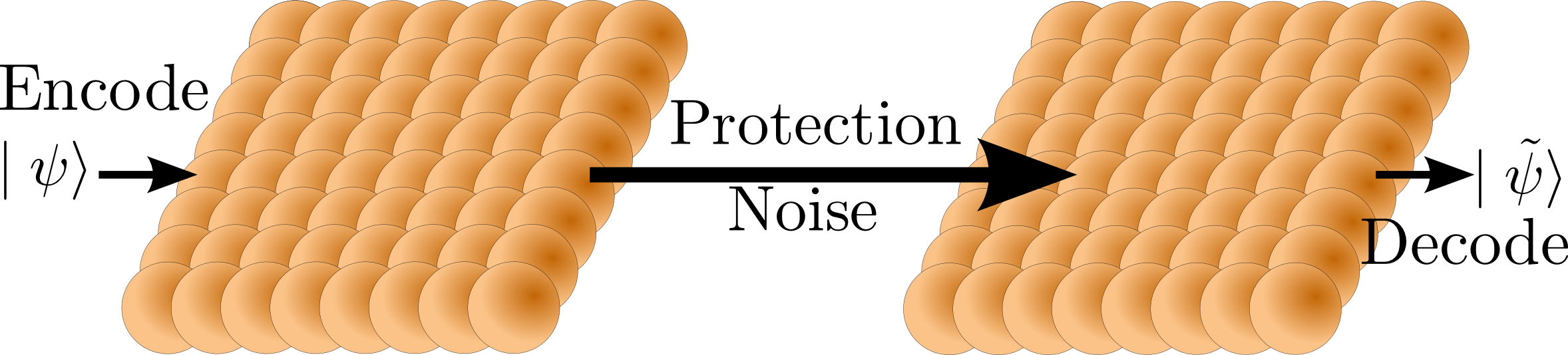}
\end{center}
\caption{ \label{fig:ManyBodyEncoding} (color online)
We assume that a piece of quantum information is encoded into a many body system.
The engineered dissipation, is then responsible for making the degrees of freedom which carry the encoded quantum information resilient against the uncontrolled noise processes taking place.
Finally, the decoding process extracts the quantum information from the collective degrees of freedom.}
\end{figure}

\section{Statement of the problem\label{sec:StatementOfproblem}}

We consider a logical qubit encoded in $N$ physical qubits, which are appropriately coupled to an
environment providing dissipation.
We describe the action of the engineered environment, as well as of the other sources of decoherence through a master equation
 \begin{equation}\label{eq:LdissLnoise}
 \dot{\rho} = \mathcal{L}_{\rm diss}(\rho) +  \mathcal{L}_{\rm noise} (\rho)
 \end{equation}
Here, $\rho$ is the density operator for the qubits, $\mathcal{L}_{\rm diss}$ the Liouvillian
describing the engineered dissipation, and $\mathcal{L}_{\rm noise}$ will denote a noise term contribution to the master equation.
This could be local depolarizing noise for instance
 \begin{equation}\label{eq:Ldep}
\mathcal{L}_{\rm noise} (\rho) = \Gamma_\epsilon \mathcal{L}_{\rm dep} (\rho)=  \Gamma_\epsilon \sum_{n =
 1}^N \frac{\mathbbm{1}_n}{2} \otimes \tmop{tr}_n (\rho) - \rho,
 \end{equation}
or any other weak local noise term.
Our goal is to show that for appropriate choices of $\mathcal{L}_{\rm diss}$
we can extract the encoded qubit reliably after a time which substantially increases with $N$.

In general, any trace preserving dissipative master equation as $\mathcal{L}_{\rm diss}$  may be write in Lindblad form \cite{lindblad_generators_1976}
\begin{equation}\label{eq:LinbladME}
\dot{\rho}=\mathcal{L}(\rho) = -i[H,\rho] + \sum_k 2L_k \rho_0 L_k^\dagger - \{ L_k^\dagger L_k \rho \}_+,
\end{equation}
consisting of a Hamiltonian term describing the unitary evolution, and a dissipative part which may be written in terms of Lindblad or jump operators $L_k$.
Furthermore, the models of engineered dissipation we propose can be seen to adhere to a more benign form
\begin{equation}\label{eq:StochasticME}
 \dot{\rho} = \sum_l \Gamma_l \left[ T_l(\rho) -\rho \right],
\end{equation}
where $T_l$ are positive trace preserving channels.
For these particular cases, the time dependent density matrix may be given an explicit stochastic expansion in the form of
\begin{equation}
 \rho(t) = e^{-\Gamma t}\sum_{n=0}^\infty  \frac{T^n \rho(0)}{n!},
\end{equation}
where $\Gamma = \sum_l \Gamma_l$ and $T(\rho)=\sum_l \Gamma_l T_l(\rho)$.
This stochastic expansion will be useful for both proofs and Monte Carlo simulations.

\section{Straightforward QECC encoding}

Here we introduce and analyze two straightforward methods of encoding a
QECC in the dissipation. 
The first one consist of coupling all the qubits with a reservoir in such a way that each application of a jump operator a whole error correction procedure takes place. 
In the second, we encode the QECC in several Lindblad terms, so that each jump correspond to an execution of a part of the QEC. 
The main purpose of this section is to show that those simple approaches do not work as one could imagine, and thus it illustrates why the design of engineered quantum memories is not a trivial task. 
Both approaches require multibody coupling to a single environment, where the number of system qubits coupled to the same {\it damped qubit} grows with $N$, the size of the memory.
In principle perturbation theory gadgets allow the engineering of such terms, provided their respective intensity decay exponentially with the number of subsystems involved.
Not withstanding, a strength increasing with $N$ would be required to make the first approach work, while in the second approach only a polynomial decrease with the number of subsystems involved would preserve functionality.
In the next section we will present a scheme which circumvents these problems, although still with the caveat that it requires non-local couplings (as it works in 4 spatial dimensions).

\subsection{Single Jump Operator}

One major obstacle to traducing the usual error correction strategies to a dissipative scenario is due to the random times at which dissipative terms enact the recovery operations.
We illustrate this problem in the case of a straightforward approach to dissipative protection.
One can always implement in the dissipative Liouvillian
a standard quantum error correction procedure  which preserves the logical qubit:
$\mathcal{L}_{\rm diss}(\rho)=\Gamma[\mathcal{R}(\rho) - \rho]$, where
$\mathcal{R}$ is a full recovery operation and $\Gamma$ adjusts the rate at which
the recovery operation is applied (imagine full correction of an $N$ qubit QECC).
Apart from the unrealistic nature of highly many--body dissipation  terms required in this construction, it is easy to see that it does not serve our purposes.
The reason can be seen by unraveling the quantum jump operators \cite{carmichael_statistical_1998}, there is a finite $N$ independent probability for more than $\frac{1}{\Gamma_\epsilon}$ time to elapse until the next recovery operation.
Such long times allow too many errors to accumulate for any QECC to recover with high fidelity.

The alternative is to have dissipation implement many independent processes instead of a single monolithic error correction procedure.
Ideally, having independent processes take care of removing independent error sets can make the accumulation of a critical fraction of errors exponentially unlikely.
The difficulty of having independent dissipation processes is that contrary to the circuit model the order of their application is not enforced in any way.
Thus, directly encoding each gate of a QECC recovery circuit into a dissipation term generally leads to a meaningless evolution.
However, we will show that in specific cases where dissipation terms commute or show some order property lending itself to rigorous analysis, the asynchronous nature is not an obstacle.

\subsection{Concatenated QECC Dissipation}

It is indeed possible to design a many-body dissipative quantum memory.
The strategy is to take the dissipation term as a sum of recovery operations occurring on different groups of qubits.
Those operations correspond to recovery of the different logical qubits at each level of a simple concatenated QECC \cite{knill_resilient_1998}.
Intuitively, one may argue that the difficulty of implementing a given dissipation term increases with the number of qubits involved.
We attempt to compensate for this difficulty by imposing that the
operator norm required for such Lindblad terms decays with a power law respect
to the number of physical qubits involved.
More specifically, we take
\begin{equation}
   \mathcal{L}_{\rm diss}(\rho) = \Gamma \sum_{l,n} \delta^{M-l}
   [\mathcal{R}_{l,n} (\rho) -\rho].
\end{equation}
Here, $l=0,1,\ldots,M-1$ denotes the level of concatenation, and $n$ further specifies
on which set of qubits the recovery operations $\mathcal{R}_{l,n}$ are applied.
In appendix \ref{sec:CCD}, we show that if the local noise rate $\Gamma_\epsilon$ is sufficiently small then initially encoded information is lost at a rate which is exponentially small
with respect to the number of qubits used (i.e. double exponentially small
with the level of concatenation $M$).
The weakness condition on the noise can be made precise by
\begin{equation}
  \Gamma_{\epsilon} < \Gamma_\epsilon^\star = \frac{\delta^2 \Gamma}{k^2} ,
\end{equation}
where $k$ is the number of physical qubits in the code to be concatenated.
Assuming the perfect 5 qubit QECC and taking the strength of many body terms
inversely proportional to the number of bodies ($\delta = 1 / 5$), a threshold
of $\Gamma_\epsilon^\star = 1.6 \times 10^{- 3} \Gamma$ is obtained for the noise rate.
When the error rate is below the error threshold,
the relaxation rate for the encoded information has an exponentially
small upper bound given by
\begin{equation}
  \tau^{-1} \leq \Gamma_\epsilon \delta^M \left(\frac{\Gamma_\epsilon}{\Gamma_\epsilon^\star}\right)^{2^M-1}
\end{equation}

The above scheme is mainly of formal interest, since the non local recovery operations encoded in the dissipative master equation require many qubits at different locations to interact with the same environment.
While the necessary scaling of such terms  needs to be polynomial for our proof to go through, the derivation of such terms based on effective many-body Hamiltonians and the dissipative gadgets we propose is expected to decay exponentially with the number of bodies involved.
Even more realistically, one would expect many-body dissipation terms to cope with many body error terms arising from imperfect implementation.
In practice, it would be desirable to find a set up where the dissipation
terms are spatially localized by considering the qubits arranged in a lattice.

\section{Local dissipative protection in 4D}

In classical systems Toom's rule \cite{Toom1974}  has been proven to be a
simple translationally invariant update rule in a 2D Periodic Boundary Condition (PBC) lattice which is capable of preserving classical information, provided that the noise contribution to the dynamic is sufficiently weak.
While we have not been able to extend this rule for quantum
protection in 2D, we will consider a quantum analog of Toom's rule for 4D.
The underlying QECC used is the 4D toric code, a stabilizer quantum error correcting
code with 6 body stabilizer generators which can be made spatially local in a 4D PBC lattice.
Dennis et al.\cite{dennis_topological_2002} proposed it as a local QECC, and the corresponding stabilizer Hamiltonian was recently rigorously proven to be {\it self--correcting} by Alicki et al. \cite{alicki_thermal_2008}.
We derive a local master equation for protecting information encoded into the 4D toric code
based on a Toom like rule introduced by Ahn \cite{ahn_extending_2004} and study its
efficiency for protecting encoded observables.
We then consider the protection process and numerically study the lifetime of information when depolarization errors are introduced extensively at a small yet constant rate.

\begin{figure}[h]
\centering
\subfigure[Lifetime Vs. error rate]{
\label{fig:LifetimeVsEpsilon}\includegraphics[width=1.0\columnwidth]{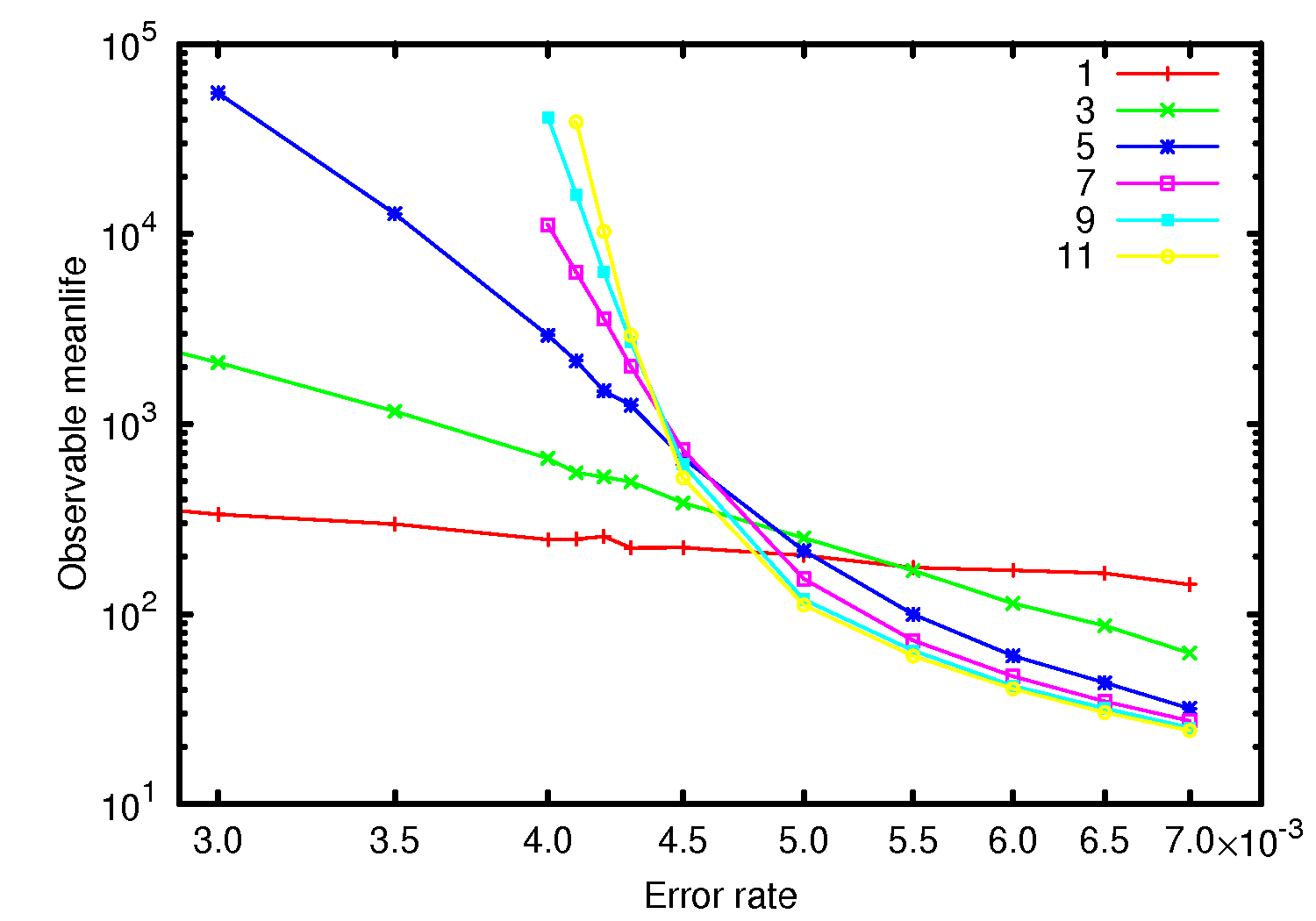}
}

\subfigure[Lifetime Vs. lattice size]{
\label{fig:LifetimeVsN}\includegraphics[width=1.0\columnwidth]{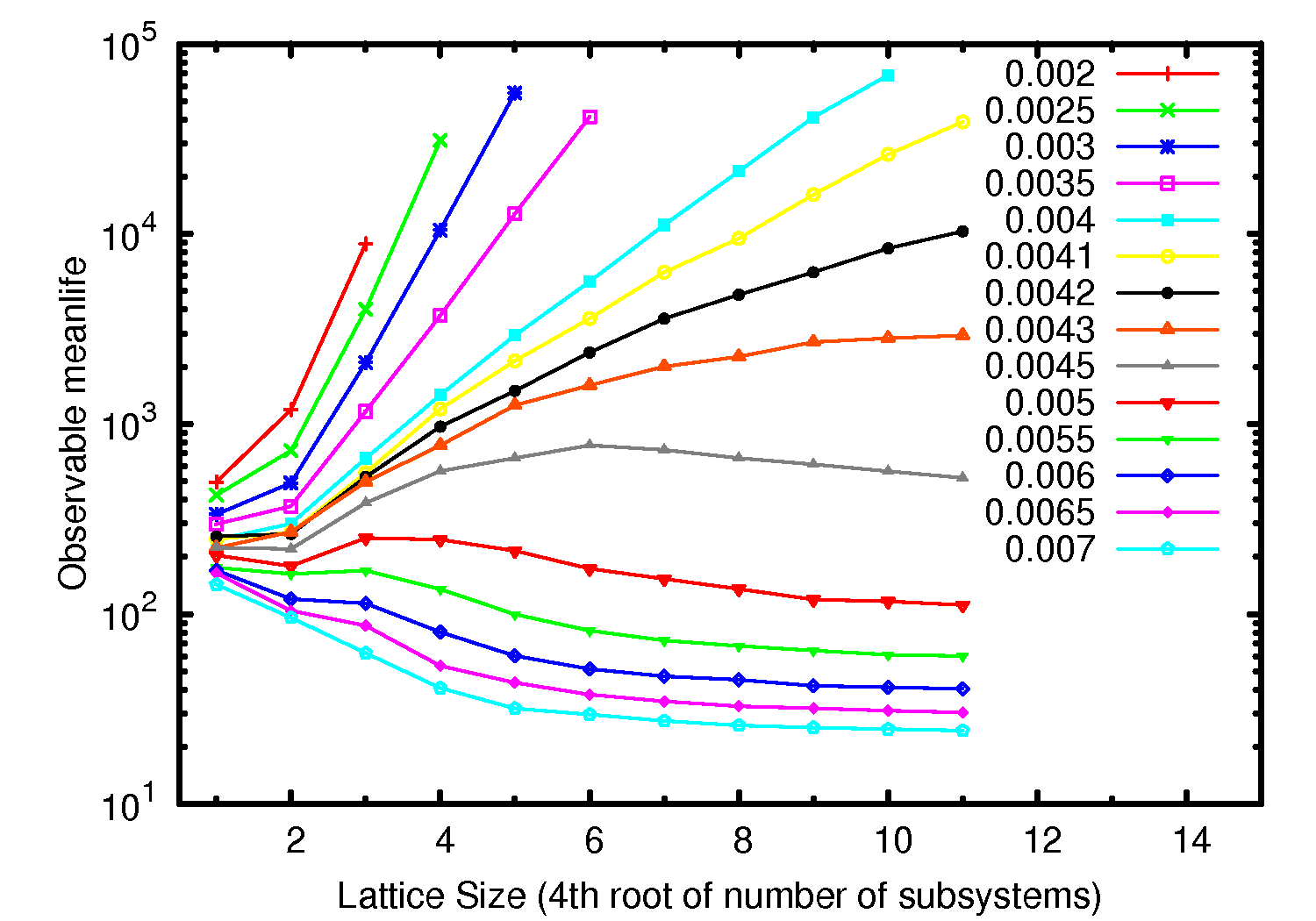}
}

\caption{ (color online)
The mean time to error for a logical observable is plotted in log scale units of $\frac{1}{\Gamma}$.
Error rates $\Gamma_\epsilon$ are provided in units of $\Gamma$.
The plots further suggests the existence of a critical value for error rates $\Gamma_\epsilon^\star \approx 0.004$.
(a) Each curve corresponds to a fixed odd value of the lattice size $N$.
The independent axis $\Gamma_\epsilon/\Gamma$ is also in log scale suggesting that for each fixed $N$ the information lifetime show an asymptotic (small $\Gamma_\epsilon$) power law dependence with $1/\Gamma_\epsilon$ with the exponent increasing for larger $N$.
(b) Each curve corresponds to a fixed value of the error rate $\Gamma_\epsilon$.
For low error rates $\Gamma_\epsilon < \Gamma_\epsilon^\star$, lifetime is seen to improve exponentially with $N$.
}
\label{fig:Lifetime4DTC}
\end{figure}

A fully rigorous description of the QECC and the local update rule used is provided in
the appendix \ref{sec:4DToom}.
For the moment it is sufficient to specify that the master equation has the form of eq. (\ref{eq:LdissLnoise}) where the specific $\mathcal{L}_{\rm diss}$ used associated to the 4D toric code will be called $\mathcal{L}_{\text{4D-TCToom}}$ and $\mathcal{L}_{\rm noise}$ is weak extensive depolarizing noise as in eq. (\ref{eq:Ldep}).
The numerical results (Fig. \ref{fig:Lifetime4DTC}) strongly support the existence of a critical error rate $\Gamma_\epsilon^\star \approx 0.004\times\Gamma$ ( where $\Gamma$ is the correction rate to be specified ) below which, the lifetime of the encoded information increases exponentially with the lattice size.

Although the results above have no obvious practical implication, they
suggest that local models may exist in spatial dimensions lower than 4
(for the search of quantum memories based on protecting Hamiltonians
in lower dimension see \cite{bacon_operator_2006,chesi_self-correcting_2009, hamma_topological_2008}).
The hope, is that even if self-correcting quantum memories fail to exist in lower dimensions, the use of engineered dissipation may still provide a solution.

\subsection{Numerical simulations}

The key feature that allows us to perform efficient
simulations of the relaxation times for logical observables, is that the terms in
$\mathcal{L}_{\text{4D-TCToom}}$ may be naturally split in two subsets, where terms in
one subset commute with terms in the other.
Thus, efficient classical Monte Carlo simulations provide unbiased estimates for expectation values and correlations for half of the stabilizers and half of the logical observables.
Throughout each simulation the relevant error corrected
logical observable was measured on a copy of the system state after every unit of simulated time.
Simulation were interrupted when a measurement outcome differing from the initial value was obtained.
For each parameter, lattice size $N$ and the depolarization rate $\Gamma_\epsilon$, a total of 1000 such runs were performed and the simulated times were averaged to obtain the relaxation time presented.
These simulations where performed on 62 AMD Opteron processors taking a total of five days to
obtain the  data presented (Fig. \ref{fig:Lifetime4DTC}).

\section{Accessible toy model}

As a a proof of principle, we now present an engineered dissipation toy model providing protection for quantum information.
One can implement the underlying ideas of dissipative quantum memories with 2D lattices at the expense of being able to correct only for dephasing noise
 \begin{equation}
 \mathcal{L}_{\rm phase} (\rho)= \Gamma^z_\epsilon \sum_{n =
 1}^N \sigma^z_n (\rho)\sigma^z_n - \rho.
 \end{equation}
Given a noise model including only one type of error (such as $\sigma^z$ phase errors) we will be able to cast a classical memory prescription into a quantum scenario.
A first step, is to define two logical observables
\begin{equation}
 Z^{EC} \equiv \bigotimes_s \sigma^z_s \qquad X^{EC} \equiv \theta\left(\sum_s \sigma^x_s\right)
\end{equation}
where $\theta$ is the Heaviside step function.
The first observable $Z^{EC}$ commutes with the noise ${\mathcal L}_{\rm phase}$ and is thus completely immune to it.
The noise can only change the value of $X^{EC}$, for the part of $\rho$ which is in the $\pm1$ eigenspace of $ \sum_s \sigma^x_s $ (i.e. states for which the absolute magnetization in the $X$ direction is minimal).
Dissipation will protect the $X^{EC}$ observable by keeping most of $\rho$ in a high $X$ magnetization subspace.
The master equation $\dot{\rho} = \mathcal{L}_{\text{NN}}(\rho)$ for nearest neighbor majority voting is written as a Liouvillian in  Lindblad form \cite{lindblad_generators_1976} as
\begin{equation}
  \mathcal{L}_{\text{NN}}(\rho) = \Gamma \sum_{<s, r,t>} L_{s, r,t} \rho L^{\dagger}_{s, r,t} -
  \frac{1}{2} \{L^{\dagger}_{s, r,t} L_{s, r,t}, \rho\}_+,
\end{equation}
where the index $s$ runs over all sites, $r\neq t$ are nearest neighbors of $s$ and the Lindblad operators are given by
\begin{equation}
  L_{s, r, t} \equiv \sigma^z_s \frac{1 - \sigma^x_s \otimes \sigma^x_r}{2}
  \frac{1 - \sigma^x_s \otimes \sigma^x_t}{2}.
\end{equation}
This is, the first factor performs a phase flip when the second and third factors (projectors)
are non zero (i.e. when site $s$ points differently than its two neighbors $r$ and $t$).
The Lindblad operators are designed such that they also commute with $Z^{EC}$ and can only change $X^{EC}$ in the portion of $\rho$ with minimal $X$ magnetization.

The stability of the $X^{EC}$ observable in such an evolution can be mapped to magnetization metastability in classical studies \cite{binder_scaling_1973, cirillo_metastability_1998}.
Restricting $r$ and $t$ to be north and east neighbors in an $N\times N$ PBC lattice, one recovers  Toom's rule \cite{Toom1974, gray_Tooms_1999} which is proven to provide an exponential survival time, even in the presence of biased errors.
However, the PBC requirement is experimentally unrealistic.

\begin{figure}[h]
\begin{center}
\includegraphics[width=1.0\columnwidth]{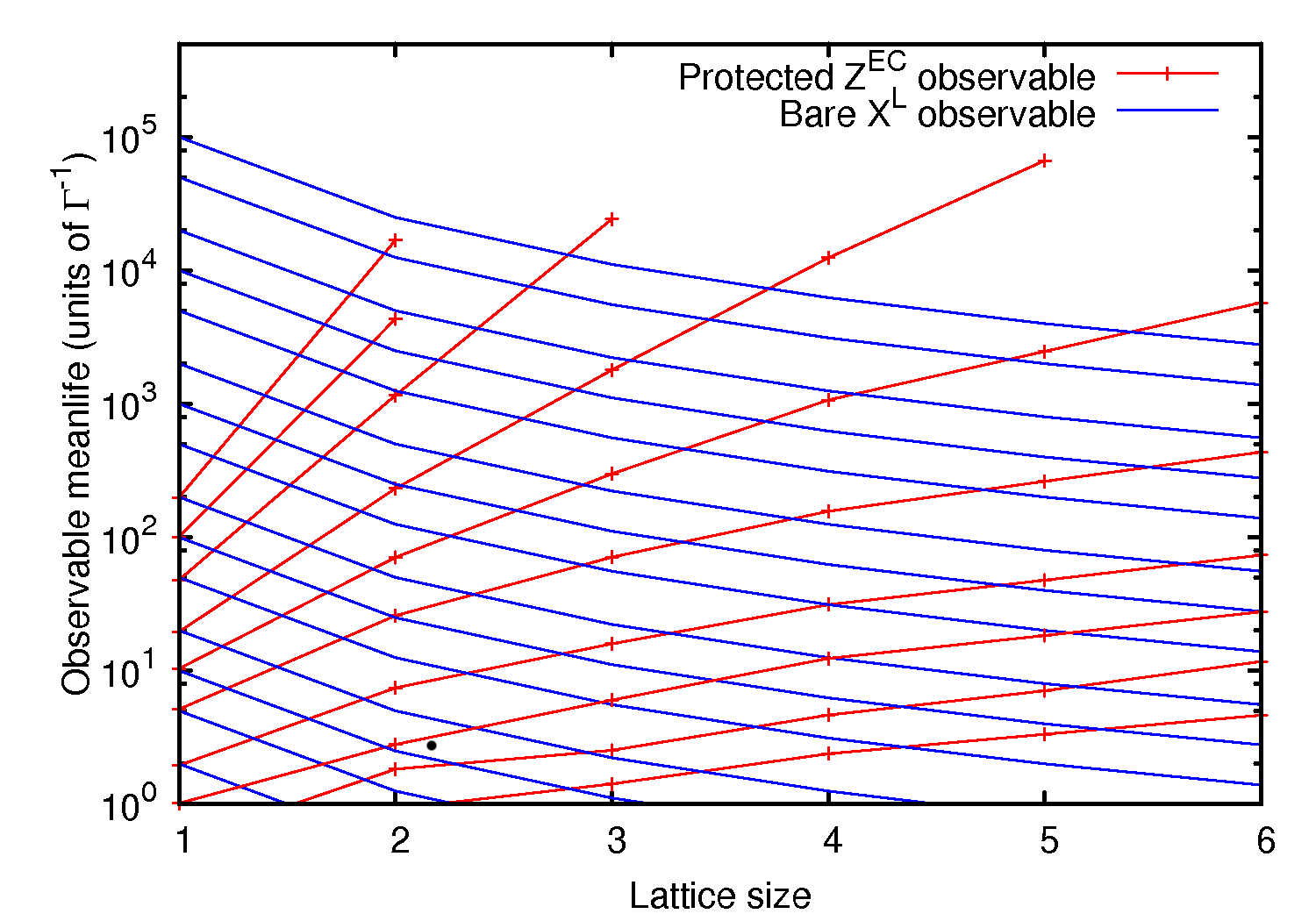}
\end{center}
\caption{ \label{fig:2DMajVote} (color online)
Relaxation time for $Z^{EC}$ (red curves) and $X^{L}$ (blue curves) in units of $\Gamma^{-1}$.
Each red curve presents the relaxation time $\tau_Z$ (numerically obtained) corresponding to one value of the relative dephasing rate $\Gamma/\Gamma_{phase}$ given by the intercept at $N=1$.
Blue curve have the functional form $\tau_X = \Gamma_{dep}^{-1}*N^{-2}$ and each corresponds to one value of $\Gamma/\Gamma_{dep}$ also given by the intercept at $N=1$.
The lifetime $\tau$ of the encoded logical qubit can be seen to be estimated by $\tau \approx min\{\tau_X, \tau_Z\}$.
Given $\Gamma/\Gamma_{dep}$ and $\Gamma/\Gamma_{phase}$, one may intersect the corresponding curves  to obtain the value of $N$ leading to the optimal qubit lifetime $\tau$.
For example, if $\Gamma_{dep}=5\times10^{-5}\Gamma$ and $\Gamma_{phase}=0.1 \Gamma$ the optimal lattice size of $4\times 4$ allows a $\times100$ increase in the quantum information relaxation time $\tau$.
A more extreme case may be seen when $\Gamma_{phase}=0.01 \Gamma$ and $\Gamma_{dep}\leq 5\times 10^{-5} \Gamma$ where a factor $\times 50$ is gained by simply using a $2\times 2$ lattice.
}
\end{figure}

We numerically consider an experimentally accessible setup which does not require periodic boundary conditions.
Physical qubits will be located on an $N\times N$ 2D square lattice sites.
The sites $r$ and $t$ are taken among all possible nearest neighbors of $s$.
The number of valid neighbor combinations are $\binom{4}{2}=6$ for inner sites $s$, $\binom{3}{2}=3$ for lattice border sites $s$ and only one combination for corner sites.
In the following plot (Fig. \ref{fig:2DMajVote}), we show how having a protective dissipation term ${\mathcal L}_{NN}$ can increase the relaxation time of $X^{EC}$, a many-body encoded observable (red).
This is in contrast to the complementary observable which does not benefit from dissipative protection.
On the contrary, given any depolarization rate, the relaxation time of $Z^{L}$ decreases with the inverse of the number of physical qubits involved (blue).

\section{Dissipative gadgets}

As we have shown, the possibility of controlled quantum dissipation opens a host of new possibilities for QIP \cite{diehl_quantum_2008, verstraete_quantum_2009, krauter_entanglement_2010}.
However, while some naturally occurring forms of dissipation may be readily exploited, it is crucial to have a systematic way of engineering arbitrary dissipative dynamics.
A way of achieving complete control over the dissipation is to be capable of engineering independent Lindblad jump operators while keeping their interference with each other weak.
For this we must assume availability of many body Hamiltonians, achievable through perturbation theory gadgets \cite{jordan_perturbative_2008, bravyi_quantum_2008} and of some naturally occurring dissipation, namely in the form of {\it damped qubits}.
We apply the \emph{approximation of independent rates of variation} \cite{cohen-tannoudji_atomphoton_1992} pg. 356 on the damped qubits which requires the bath correlation time for the damping process to be much shorter than the inverse of any coupling constant in the system.
Coupling to these {\it damped qubits} can thus be seen as a resource in the design of quantum dynamics, analogous to freshly initializing qubits in quantum circuits.

Coupling the system to a {\it damped qubit} ancillary degree of freedom was proposed as a possible path to engineer arbitrary effective dissipative dynamics \cite{verstraete_quantum_2009}.
More specifically, the Hamiltonian coupling $H = \omega (L \otimes \sigma^+ + L^{\dagger} \otimes \sigma^-)$ to an ancilla with damping rate $\gamma$ leads to an effective dissipative dynamics of the system corresponding to the Lindblad operator $\omega\sqrt{2/\gamma}L$.
Here $\sigma^- = \ket{0}\bra{1}$ and $\sigma^+ = \ket{1}\bra{0}$.

In order to use these dissipation gadgets as basic building blocks in more complex scenarios, it is essential to make explicit possible limitations and restrictions of the implemented dissipation.
In appendix \ref{sec:AdiabaticElimination} we provide a detailed derivation of the effective system dynamics which makes three main contributions to our understanding of dissipative gadgets.
Firstly, while the usual approach of adiabatic elimination obtains an effective dynamics in terms of a coarse grained time, our derivation shows that excluding a short initial transient period, this temporal coarse graining is not necessary.
Secondly, we provide explicit bounds on the deviation from the desired state and instantaneous dynamics which are accompanied by a smallness prefactor $(\omega/\gamma)^2$.
Finally, we include an independent internal dynamic for the system, and show, that the resulting effective dissipation carries through essentially unaffected provided the strength of the internal dynamics is sufficiently weak.
While the last point already suggests that the extensive application of local dissipation gadgets should be well behaved, a fully rigorous analysis is beyond the scope of this article.

\section{Conclusions and perspectives}
We have introduced engineered dissipation as a tool to protect against general quantum noise and proposed examples  providing protection from local noise.
In the case of concatenated code dissipation, we prove that information can be made resilient against any strictly local noise.
Numerical simulations with depolarizing noise strongly suggest dissipative protection may be made spatially local in 4D.
For purely dephasing noise we propose a dissipative protection scheme local in 2D.
Proof of principle experiments could be realized with trapped ions, or atoms in optical lattices.

A self-correcting thermalization scheme associated to the 4D toric code Hamiltonian can provide encoded quantum information similar protection against depolarizing noise.
In this sense, we have not illustrated the advantage of engineered dissipation.
While the approach we have taken with the 4D TC is analogous to Toom's 2D
update rule for classical information the thermalization of the 4D toric code can be seen as analogous to thermalization of the 2D Ising model respect to unbiased noise.
However, stretching such parallelism with the classical problem suggests that engineered dissipation may be strictly more powerful and that it may be possible to engineer a 2D local dissipation mechanism capable of protecting quantum information.
Indeed, while in 1D there can not be a self-correcting classical memory, a 1D local dissipative master equation due to G\'acs \cite{gacs_reliable_2001}  is proven to provide increased classical information lifetime with the chain size.
Inspired by G\'acs' construction, Harrington \cite{harrington_analysis_2004} has proposed a local  quantum error correction scheme in 2D capable of protecting against quantum errors.
To make this into a dissipative scheme, the requirements of a) a global synchronization clock, b) logarithmically increasing local storage space c) error free evolution of classical information, all need to be relaxed.
Whether these assumptions can be relaxed, or other schemes in 2D or 3D exist are important questions that may dictate the fate of the practical applications for dissipative quantum memories.

\section*{Acknowledgments}

This work was supported by the European Union project QUEVADIS and DFG FG635.
We thank Frank Verstate, Norbert Schuch, and Alastair Kay for helpful discussions as well as the anonymous referee for useful comments.

\bibliography{QuantumMemorybyDissipation}

\section{Adiabatic elimination of ancilla\label{sec:AdiabaticElimination}}

In this section, we prove that Master equations with arbitrary Lindblad operators may be engineered to high accuracy by coupling the system to ancillary resource qubits which are themselves being cooled.
The basic idea is to extend the system with an additional binary degree (spin 1/2) of freedom
per Lindblad operator $L$ to be implemented. 
These degrees of freedom are further assumed to be strongly dissipatively driven with a rate $\gamma$ into a $\ket{0}\bra{0}$ ground state.
We will show that a target dissipative evolution composed of a single Lindblad jump operator
\begin{equation}
 \mathcal{L}_{\tmop{target}}(\rho) = L\rho L^\dagger - \frac{1}{2}\left\{L^\dagger L, \rho \right\}_+,
\end{equation}
may be implemented within a small error margin.
The technique used for the proof follows the adiabatic elimination of the excited ancilla subspace in spirit, but takes into account corrections in order to provide rigorous bounds on the deviations from the intended evolution.

Our derivation starts by assuming that the full dynamics of the system can be written as
\begin{equation}
  \dot{\rho} = - i [H, \rho] + 2\gamma \sigma^- \rho \sigma^+ - \gamma
  \left\{ \sigma^+ \sigma^-, \rho \right\}_+ + \gamma{\mathcal L}_{\tmop{sys}}(\rho)
\end{equation}
where $\sigma^+=\ket{1}\bra{0}_A$ and $\sigma^+=\ket{1}\bra{0}_A$ are raising and lowering operators on the ancilla qubit and the Hamiltonian $H$ couples the system to the ancilla
\begin{equation}
  H = \omega (L \otimes \sigma^+ + L^{\dagger} \otimes \sigma^-)
\end{equation}
and ${\mathcal L}_{\tmop{sys}}$ is an additional evolution term with no effect on the ancillas.
Here, the assumption that is implicitly being made, is that we may independently sum the interaction Hamiltonian $H$ to the dissipative dynamics on both the system and the ancilla.
In the case of the ancilla decay this is the \emph{approximation of independent rates of variation} \cite{cohen-tannoudji_atomphoton_1992} pg. 356, which assumes correlation times for the reservoir responsible for spontaneous decay to be much shorter than any other relevant time in the system.
An important example where this approximation holds to a great degree of accuracy is for two level atoms at optical frequencies, where the autocorrelation time of the coupled vacuum fluctuations can be as much as ten orders of magnitude shorter than the inverse of any of the other coupling constants.
Since our derivation for the weak system Liouvillian does not require temporal coarse graining, the successively incorporation of Hamiltonian interactions rigorously leads to the additive appearance of the desired Liouville terms up to leading order.
Assuming $\epsilon = \omega/\gamma \ll 1$ we can rescale to a unitless time by incorporating a factor $\gamma$ leading to the following differential equations for the reduced density matrices.
\begin{align}
  \dot{\rho}_{00} := \bra{0}\dot{\rho}\ket{0} &= 2 \rho_{11} - i \epsilon L^{\dagger} \rho_{10} + i\epsilon \rho_{01} L  + {\mathcal L}_{\tmop{sys}}(\rho_{00})\\
  \dot{\rho}_{01} :=\bra{0}\dot{\rho}\ket{1}&= - \rho_{01} + i \epsilon \rho_{00} L^{\dagger} - i \epsilon L^{\dagger} \rho_{11} + {\mathcal L}_{\tmop{sys}}(\rho_{01})\label{eq:rho01dot}\\  
  \dot{\rho}_{11} :=\bra{1}\dot{\rho}\ket{1} &= - 2 \rho_{11} - i \epsilon L \rho_{01} + i \epsilon
  \rho_{10} L^{\dagger} + {\mathcal L}_{\tmop{sys}}(\rho_{11})
\end{align}
From here, we may obtain the integral forms
\begin{align}
  \rho_{01} (\tau)  = & e^{- \tau} \rho_{01} (0) +  \int_0^{\tau} e^{- t'} {\mathcal L}_{\tmop{sys}}[\rho_{01}(\tau-t')] d t'\label{eq:rho01}\\
  +& i \epsilon \int_0^{\tau} e^{- t'} [\rho_{00} (\tau - t') L^{\dagger} - L^{\dagger} \rho_{11} (\tau - t')] d t' \nonumber\\
  \rho_{11} (\tau)  = & e^{- 2 \tau} \rho_{11} (0) +  \int_0^{\tau} e^{- 2t'} {\mathcal L}_{\tmop{sys}}[\rho_{11}(\tau-t')] d t'\\
  -& i \epsilon \int_0^{\tau}
   e^{- 2 t'} [L \rho_{01} (\tau - t') - \rho_{10} (\tau - t') L^{\dagger}] d t' \nonumber
\end{align}
Assuming the initial conditions $\rho_{01} (0) = \rho_{11} (0) = 0$, that $\|L\|\leq 1$ and $\|{\mathcal L}_{\tmop{sys}}\|\leq E\epsilon^2$, and using that $\|\rho_{00}\|+\|\rho_{11}\|\leq 1$  we may bound
\begin{align}
  \| \rho_{01} (\tau)\| \leq  \tilde{\epsilon}  \quad \text{and} \quad
  \| \rho_{11} (\tau)\| \leq \tilde{\epsilon}^2,
\end{align}
with $\tilde\epsilon=\frac{\epsilon}{1-E\epsilon^2}$.
It is now straightforward to bound $\| \dot{\rho}_{00} (\tau)\| \leq (4+E) \tilde{\epsilon}^2$.
We may now concentrate on tighter bounds composed of higher order terms in $\epsilon$ but also, of exponentially decaying terms.
A first step to do this is to perform  integraton by parts; on eq. (\ref{eq:rho01}) one obtains
\begin{align}
  \rho_{01} (\tau) =& i \epsilon \rho_{00} (\tau) L^{\dagger} - i \epsilon
  e^{- \tau} \rho_{00} (0) L^{\dagger} \label{eq:rho01tight}\\ 
   -& i \epsilon \int_0^{\tau} e^{- t'}
  \left[ \dot{\rho}_{00} (\tau - t') L^{\dagger} + L^{\dagger} \rho_{11} (\tau
  - t') \right] d t' \nonumber\\
  +&  \int_0^{\tau} e^{- t'} {\mathcal L}_{\tmop{sys}}[\rho_{01}(\tau-t')] d t' .\nonumber
\end{align}
In the case of $\rho_{11}$ we straightforwardly obtain
\begin{align}
   \rho_{11} (\tau)  = & -\frac{i \epsilon}{2} L\rho_{01} (\tau)\\
   +& \frac{i \epsilon}{2} \int_0^{\tau} e^{- 2 t'} L\dot{\rho}_{01} (\tau - t')   d t' + h.c. \nonumber \\
   +&  \int_0^{\tau} e^{- 2t'} {\mathcal L}_{\tmop{sys}}[\rho_{11}(\tau-t')] d t' \nonumber
\end{align}
This expression may be massaged into a form which may be more readily bounded.
The steps involved include, expanding $\dot{\rho}_{01}$ according to eq. (\ref{eq:rho01dot}), then expanding appearances of $\rho_{01}$ according to eq. (\ref{eq:rho01tight}) and finally integrating numerical factors and grouping terms.
After such manipulation, one reaches the expression
\begin{align}
 \rho_{11} (\tau) = & \frac{\epsilon^2}{2} \biggl[ L \rho_{00} (\tau) L^{\dagger} - e^{-
  \tau} (2 - e^{- \tau}) L \rho_{00} (0) L^{\dagger} \label{eq:rho11tight}\\
  -& \int_0^{\tau} e^{- t'}(2-e^{-t'})  L \dot{\rho}_{00} (\tau - t') L^{\dagger} d t' \nonumber\\
  -& \int_0^{\tau} e^{- t'}(2 - 2e^{-t'}) L L^{\dagger}\rho_{11} (\tau - t')  d t' \biggr] + h.c.\nonumber \\
  +& \frac{i \epsilon}{2} \int_0^{\tau} e^{- 2 t'} L{\mathcal L}_{\tmop{sys}}[\rho_{01}(\tau-t')] d t' + h.c. \nonumber \\
  +&  \int_0^{\tau} e^{- 2t'} {\mathcal L}_{\tmop{sys}}[\rho_{11}(\tau-t')] d t' \nonumber
  \end{align}
Using eqs. (\ref{eq:rho01tight}) and (\ref{eq:rho11tight}), one may prove the following higher order bounds
\begin{align}
  \| \rho_{01} - i \epsilon \rho_{00} L^{\dagger} \|
   &\leq  (2E+5) \tilde{\epsilon}^3 + \epsilon e^{- \tau} \\
  \| \rho_{11} - \epsilon^2 L \rho_{00} L^{\dagger} \|  &\leq 
   (3E+7) \tilde{\epsilon}^4 + 2\epsilon^2 e^{- \tau},
\end{align}
Inserting these bounds into the definition of $\dot{\rho}_{00}$ 
we may bound deviation from the target evolution by
\begin{equation}
   \| \dot{\rho}_{00} - 2 \epsilon^2 \mathcal{L}_{\tmop{target}}
  (\rho_{00}) - \mathcal{L}_{\tmop{sys}}
  (\rho_{00}) \| \leq  (10E+24) \tilde{\epsilon}^4 + 4 \epsilon^2
  e^{- \tau} \\
\end{equation}
After a short transient time of the order $\frac{1}{\gamma}\rm{log}\frac{1}{\epsilon}$, the exponential term can be neglected.
Furthermore, note that the internal system dynamics $\mathcal{L}_{\tmop{sys}}$ may be time dependent and thus encode correlations of different components of the system in its time dependence.

\section{4D Toric code \label{sec:4DToom}}

\subsection{The 4D Toric code as a stabilizer code}
We will now provide an informal description of the 4D toric code.
For every vertex of an $N \times N \times N \times N$ lattice, there
are 6 orientations of faces on which physical qubits are located.
Thus, the $6\times N^4$ physical qubits are arranged on the 2D faces of a 4D PBC lattice.
We can now introduce an over-complete set of local stabilizer generators for the code, half of which correspond to 1D edges, the other half corresponding to 3D cubes.
For each 1D edge, there is a tensor product operator $Z^{\otimes 6}$, the product of $Z$
operators acting on the six 2D faces to which this edge belongs.
Dual to this, for each 3D cube, there is a tensor product operator $X^{\otimes 6}$, the product of
$X$ Pauli operators over the six 2D faces of the cube.
Two edge and cube stabilizers overlap iff the edge is an edge of the cube, and
then their overlap will be in exactly two faces.
Thus all stabilizer generators are seen to commute.

\subsection{Logical degrees of freedom}
Counting of the remaining degrees of freedom additional to the stabilizer
syndrome obtained is not as straightforward as for the 2D toric code, where
every syndrome with an even number of anyons was possible.
In the 4D case, the required condition is that the set of unsatisfied stabilizers is only allowed
to be a combination of closed loops (in the lattice and dual lattice respectively).
However, one can explicitly construct six pairs of anticommuting logical operators which commute with all stabilizer terms, one pair for each of the six possible plane orientations.
From each pair, one operator is a full plane of $X$ rotations along a full plane wrapping around
the grid in one of the six possible orientations. The second operator from
each pair consists of a dual plane of $Z$ operators arranged along the
perpendicular plane orientation.
Although analogous to the logical operators on the 2D toric code, this image probably stretches our 2D or at most 3D imagination.
Thus, to obtain an intuition about this construction it is convenient to provide formal expressions which one may operate with.

\subsection{4D PBC lattice notation}
Each vertex of the 4D periodic lattice can be identified by a four component vector
$\vec{v} = v_0,v_1,v_2,v_3 \in {\mathbbm{Z}}_N^4$.
For each vertex $\vec{v}$, there are four edges $\hat{e}$, six faces $\hat{p}$ and four cubes $\hat{c}$ having the vertex as a lower corner.
These orientations may be described by four component binary vectors
\begin{equation}
  \hat{e},\hat{p},\hat{c} \in \{(v_0, v_1, v_2, v_3) \quad|\quad v_i \in \{0,1\}\},
\end{equation}
with edge $\hat{e}$, face $\hat{p}$, or cube $\hat{c}$ orientations satisfying the additional condition $\sum_{i=0}^3 v_i$ equal to $1$, $2$ or $3$ respectively.
Each physical qubit can be identified with a tuple $\vec{v},\hat{p}$, where $\hat{p}$ identifies the plane orientation and $\vec{v}$ its lower side corner.
The $Z$ type edge stabilizers $E_{\vec{v},\hat{e}}$ are given by
\begin{equation}
  E_{\vec{v}, \hat{e}} = \bigotimes_{\hat{e} \subset \hat{p}} Z_{\vec{v}, \hat{p}} \otimes Z_{\vec{v} - \hat{p}+\hat{e}, \hat{p}},
\end{equation}
with six participating physical qubits.
Finally, the $X$ type cube stabilizer $C_{\vec{v}, \hat{c}}$ are given by
\begin{equation}
  C_{\vec{v}, \hat{c}} = \bigotimes_{\hat{p} \subset \hat{c}} X_{\vec{v},
  \hat{p}} \otimes X_{\vec{v} + \hat{c}-\hat{p}, \hat{p}},
\end{equation}
also with six participating physical qubits.

We will now describe a set of logical operators commuting with all stabilizers
which will be used to encode information in absence of errors. There is one
pair of such anticommuting logical operators for each plane orientation
$\hat{p}$ and they are given by
\begin{equation}
  X^{L}_{\hat{p}} = \bigotimes_{n, m = 1}^N X_{n \hat{e}_1
  + m \hat{e}_2, \hat{p}} \hspace{2em} Z^{L}_{\hat{p}} =
  \bigotimes_{n, m = 1}^N Z_{n \hat{e}_3 + m \hat{e}_4, \hat{p}},
\end{equation}
with $\hat{e}_1 + \hat{e}_2 \equiv \hat{p}$  and $\hat{e}_3 + \hat{e}_4  \equiv \hat{p}^\perp$.
It is easy to see that according to this definition, the two logical operator
$X^{L}_{\hat{p}}$ and $Z^{L}_{\hat{p}}$ anticommute, as
they coincide only at qubit $( \vec{0}, \hat{p})$.
One can further verify that such operators commute with the complete set of stabilizers.
Finally, it is not hard to see, that if one assumes the state to be in the code
subspace (i.e. +1 eigenstate to all stabilizers), then any homologically equivalent surfaces results in equivalent definition for the operators.

\subsection{4D Quantum Toom's rule}
We now define a local update rule which will later be used in two ways, first
as a dissipation mechanism capable of keeping errors from accumulating too badly, second as the basic component of an information recovery procedure permitting removal of all errors to allow information read-out.
The update rule is analogous to Toom's rule for classical information stored in a 2D lattice.
While the prescription of Toom's rule is to flip a bit if it is different to
both its two lower side neighbors, the prescription in 4D will be to $X$
``flip'' a qubit if both its neighboring lower side $Z$ edge stabilizers are
not satisfied, but also to $Z$ ``flip'' a qubit if both its lower side $X$
cube stabilizers are not satisfied.
This is, a local rotation may be performed depending on neighboring stabilizer state.
This is in complete analogy to an interpretation of Toom's rule in terms of local stabilizers.
One property that permits analytic and numerical analysis of such a scheme is the decoupling of
recovery for $X$ and $Z$ logical operators.

For each qubit $( \vec{v}, \hat{p})$, we can write the super-operator
describing the quantum jump implementing the update rule as
\begin{equation}\label{eq:4DToomJump}
  {\mathcal R}_{\vec{v}, \hat{p}}^{Z} (\rho) = Z_{\vec{v}, \hat{p}} P^{X}_{\vec{v},
  \hat{p}} \rho P^{X}_{\vec{v}, \hat{p}} Z_{\vec{v}, \hat{p}} + P^{X\perp}_{\vec{v}, \hat{p}} \rho P^{X\perp}_{\vec{v}, \hat{p}}
\end{equation}
where $P^{X}_{\vec{v}, \hat{p}}$ is the projector onto the subspace where a
$Z$ flip should be performed on qubit $(\vec{v}, \hat{p})$ and $P^{X\perp}_{\vec{v}, \hat{p}}$ the orthogonal subspace.
Assuming $\hat{p}=\hat{e}_1+\hat{e}_2$ the projector may be defined as
\begin{equation}
  P^{X}_{\vec{v}, \hat{p}} = \frac{1}{4} (1 - E_{\vec{v},\hat{e}_1}) (1 - E_{\vec{v}, \hat{e}_2}) .
\end{equation}
Analogously, one may define an update rule ${\mathcal R}_{\vec{v}, \hat{p}}^{X}$ which in a similar way, introduces an $X$ ``flip'' depending on the corresponding projectors
$P^{Z}_{\vec{v}, \hat{p}}$ in terms of $Z$ type stabilizers.

\subsection{Full recovery and error corrected operators}
\begin{figure}[h!]
\includegraphics[width=1.0\columnwidth]{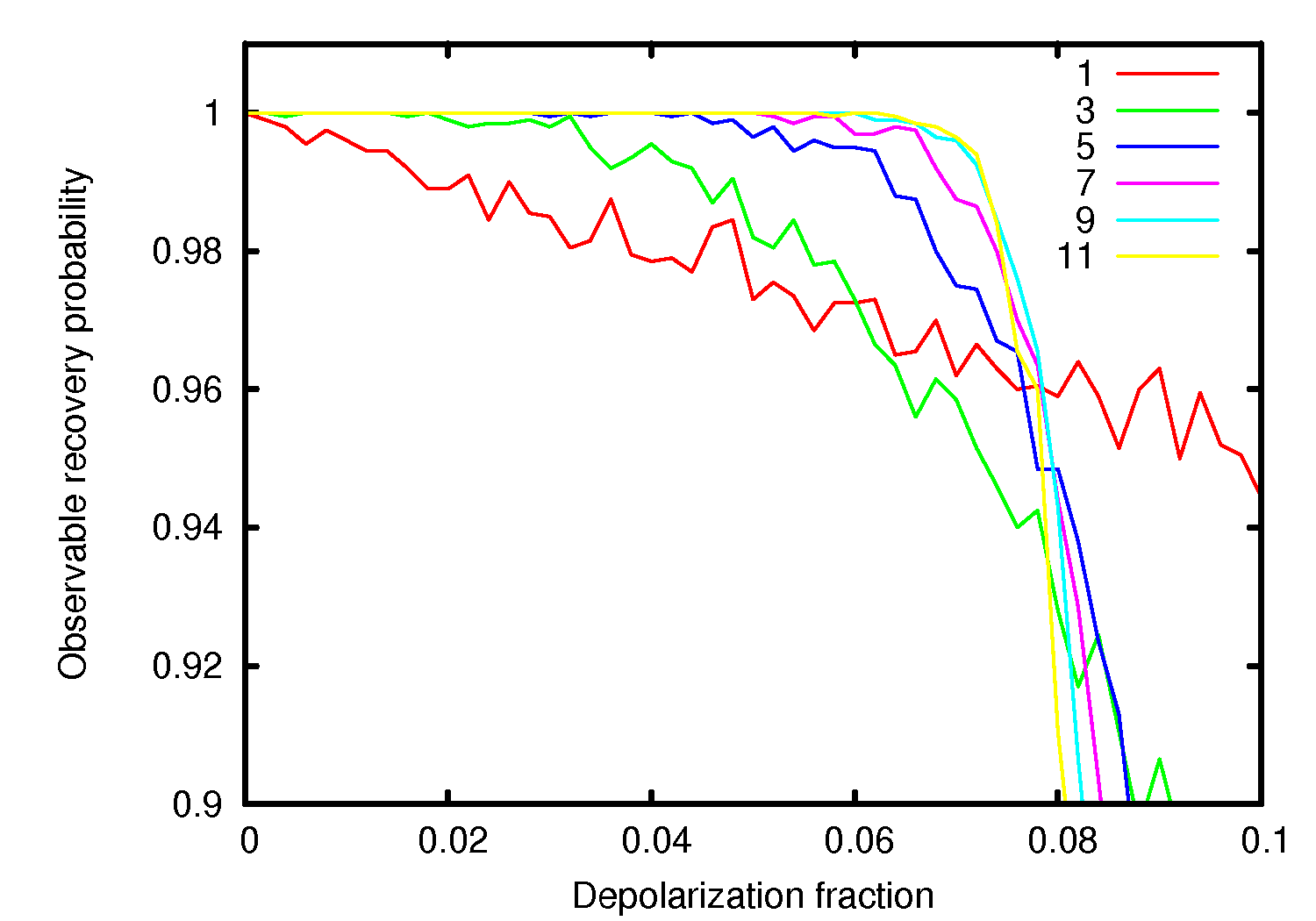}
\caption{ \label{fig:StaticQECC} (color online)
Recovery probability of an encoded observable in the 4D toric code is plotted as a function of depolarization probability per qubit.
Odd lattices sizes from 1 to 11 are represented in the different curves and suggest a critical depolarization probability of approximately $7.5\%$.
}
\end{figure}

The superoperators ${\mathcal R}_{\vec{v}, \hat{p}}^{Z}$ and ${\mathcal R}_{\vec{v}', \hat{p}'}^{X}$
always commute.
Only recovery operators of the same kind may lack commutation
when considering neighboring plaquetes.
In particular, to define a full recovery operation $\mathcal{R}$ in terms of these local recovery update rules, it is necessary to unambiguously specify an order of application.
Indeed, in our simulation code, a sweep through the lattice is taken as this order and we observe a good performance in recovering the originally encoded observables (Fig. \ref{fig:StaticQECC}).
Once the recovery operation $\mathcal{R}$ is unambiguously specified,
it is possible to define robust logical observables
$X^{EC}_{\hat{p}}$ and $Z^{EC}_{\hat{p}}$ such that
\begin{equation}
  \tmop{tr} (Z^{EC}_{\hat{p}} \rho) = \tmop{tr}
  (Z^{L}_{\hat{p}} \mathcal{R}\rho) \hspace{2em} \tmop{tr}
  (X^{EC}_{\hat{p}} \rho) = \tmop{tr} (X^{L}_{\hat{p}}
  \mathcal{R}\rho).
\end{equation}
Or more compactly
\begin{equation}
 Z^{EC}_{\hat{p}} = \bar{\mathcal{R}} \left(
Z^{L}_{\hat{p}} \right)\text{ and }X^{EC}_{\hat{p}} =
\bar{\mathcal{R}} \left( X^{L}_{\hat{p}} \right).
\end{equation}
Thus, error corrected logical observables (super-index $EC$), provide a robust result when
evaluated on a state with sufficiently few errors and coincide with logical operators on the error-free subspace.

\subsection{Master equation}
We study a master equation including a locally depolarizing noise term of
strength $\Gamma_\epsilon$, and the proposed Lindblad terms intended
to avoid error clusters from growing.
The simulated master equation may be written as
\begin{equation}
  \dot{\rho} = \mathcal{L}\rho = \Gamma \mathcal{L}_{4DToom}\rho + \Gamma_\epsilon \mathcal{L}_{dep}\rho
\end{equation}
where the dissipative protection $\mathcal{L}_{4DToom}$ is given by
\begin{equation}
\begin{split}
 \mathcal{L}_{4DToom}\rho = \sum_{\vec{v}, \hat{p}}&
  L^{X}_{\vec{v}, \hat{p}} \rho  L^{X\dagger}_{\vec{v}, \hat{p}} - \frac{1}{2}\left\{L^{X\dagger}_{\vec{v}, \hat{p}}L^{X}_{\vec{v}, \hat{p}}, \rho\right\}_+ \\
  + & L^{Z}_{\vec{v}, \hat{p}} \rho  L^{Z\dagger}_{\vec{v}, \hat{p}} - \frac{1}{2}\left\{L^{Z\dagger}_{\vec{v}, \hat{p}}L^{Z}_{\vec{v}, \hat{p}}, \rho\right\}_+.
\end{split}
 \end{equation}
The protecting Lindblad operators are
\begin{equation}
  L^{Z}_{\vec{v}, \hat{p}} = Z_{\vec{v}, \hat{p}} P^{X}_{\vec{v}, \hat{p}}
  \quad L^{X}_{\vec{v}, \hat{p}} = X_{\vec{v}, \hat{p}}
  P^{Z}_{\vec{v}, \hat{p}},
\end{equation}
corresponding to the Toom like quantum jump superoperators ${\mathcal R}_{\vec{v}, \hat{p}}^{\{X,Z\}}$ introduced in Eq. \ref{eq:4DToomJump}.
We perform numerical experiments to determine the relaxation time for logical observables (i.e. $\tmop{tr}[X^{EC}\rho(t)]\equiv\tmop{tr}[X^L {\mathcal R}\rho(t)]$.)
Evolutions are taken to start in a code state with an unambiguous $X^L$ or $Z^L$ logical value and consistency of the error corrected logical observables are checked regularly in time.
The mean time to the first change in the value observed for $X^{EC}$ or $Z^{EC}$ is taken as an estimator of the relaxation time.

\subsection{Numerical considerations}
Evolution under this master equation can be numerically simulated efficiently
for a commuting set of observables such as the edge stabilizers and a
commuting set of logical observables.
This means that a classical Monte Carlo simulation is enough to study the probability of obtaining the correct outcome when measuring distinct logical observables which were initially well defined.
The results of such simulations are presented in (Fig. \ref{fig:Lifetime4DTC}) for different sizes of the lattice grid up to 11 and different values for $\Gamma_\epsilon$.
These suggest a critical value for the noise rate $\Gamma_\epsilon^\star \approx 0.004$, below which arbitrarily long relaxation times may be achieved by increasing the lattice size.
Given that below threshold error rates, the information lifetime seems
to grow exponentially with the lattice size, and that the simulation time per
unit time is also proportional to the fourth power of the lattice size, it is
numerically costly to extend our evidence to larger lattices.

\subsection{Definition of efficient recovery ${\mathcal R}$}
To check whether the encoded observable is still recoverable at time $t$, we apply a correction super-operator ${\mathcal R}$ on $\rho(t)$.
The definition of ${\mathcal R}$ consists of sequentially applying the local jump superoperators ${\mathcal R}_{\vec{v}, \hat{p}}^{\{X,Z\}}$ in a sweeping order.
This shows a high performance for removing all error domains (i.e. it presents a numerical threshold to depolarizing noise on up to $\approx 7.5\%$ of the qubits as shown in (Fig. \ref{fig:StaticQECC})
.)
Furthermore, computer simulations of ${\mathcal R}$ are efficient, requiring a minimal amount of $O(N^4)$ operations.
This is important since the ${\mathcal R}\rho(t)$ must be checked every unit time to obtain an estimate of the relaxation times of error corrected logical observables.

\section{Concatenated-code dissipation \label{sec:CCD}}

Paz and Zurek \cite{paz_continuous_1998} presented the first studies of protecting quantum information through the use of a continuous dissipative process.
They introduce a general master equation form for the class of stabilizer QECC and analyze their performance in some simple cases.
In this section, we will adapt their construction and propose master equations for concatenated QECC which are provably robust against sufficiently weak local noise.

For stabilizer QECC, the recovery super-operator ${\mathcal R}$ can be written in Kraus form as
\begin{equation}\label{eq:RecoveryOperator}
 {\mathcal R}(\rho) = \sum_{(m)} R^{(m)} P^{(m)} \rho P^{(m)\dagger} R^{(m)\dagger} ,
\end{equation}
where $P^{(m)} $ are projectors onto orthogonal syndrome subspaces with $\sum_{(m)} P^{(m)} = {\mathbbm 1}$ and $R^{(m)}$ are unitary recovery operators of tensor product Pauli form.
The operators $L^{(m)} = R^{(m)} P^{(m)}$ can be interpreted as Lindblad operators to give way to a protecting master equation.
However, this approach can not provide more than a constant improvement in the relaxation time for logical observables.
It can be seen that given an error rate and a correction rate, there is an upper bound on the relaxation time of the logical encoded bit which is independent of the code and the number of physical subsystems it uses.

We propose extending this master equation model to one which allows performing many such recovery operations in parallel.
In the case of concatenated QECC these will correspond to error correction at the different levels of concatenation.
Recovery operations at the same level of concatenation act independently of each other as they involve disjoint subsets of physical qubits.
Most of the work goes into designing recovery operations at different levels of concatenation which do not interfere undesirably (i.e. commute) and proving that they provide a similar protection from local errors to the one achieved by concatenated QECC in the circuit model.

We will define a dissipative concatenated quantum memory based on a $[[k, 3, 1]]$ QECC.
An $M$-level encoding will thus make use of $k^M$ physical qubits.
A labeling for each physical qubit may be given in terms of an $M$ component vector $\vec{v}\in {\mathbb Z}_k^M$ (i.e. with each component going from $1$ to $k$.
Partial vectors $\vec{v}$ with $M-l$ components will identify mutually disjoint blocks of $k^l$ physical qubits.
Thus if $\vec{v}$ denotes a particular set of $k^l$ physical qubit, then the vector $v_0 : \vec{v}$, with one additional component $v_0$ and identifies a sub-block of $k^{l-1}$ physical qubits.
The number of components of a vector $\vec{v}$ will be denoted by $\mid \vec{v} \mid$, with $\emptyset$ being the unique zero component vector.

A stabilizer QECC on $k$ qubits can be characterized by the definition of the Stabilizers $S^{(j)}$, the projectors onto syndrome subspaces $P^{(j)}$ and the corresponding error recovery operators $R^{(j)}$, the logical operators $X^L, Y^L, Z^L$, the recovery super-operator ${\mathcal R}$ and the error corrected Pauli observables $X^{EC}, Y^{EC}, Z^{EC}$.
It is instructive to present the definition of these objects for a simple QECC making it easier to latter provide the recursive definitions required for the concatenated QECCs.
These definitions are given by
\begin{equation}
 \begin{split}
 S^{(j)} & =  s1^{(j)}\otimes s2^{(j)} \otimes \ldots \otimes sk^{(j)} \\
 P^{(j)} & =  \sum \alpha_{i,j} S^{(i)} \\
 R^{(j)} & =  r1^{(j)}\otimes r2^{(j)} \otimes \ldots \otimes rk^{(j)} \\
 {\mathcal R}(\rho) & = \sum R^{(k)}P^{(k)} \rho P^{(k)}R^{(k)} \\
 \sigma^L & =  \sigma1 \otimes \sigma2 \otimes \ldots \otimes \sigma k \\
 \sigma^{EC} & =  \overline{\mathcal R}(\sigma^L).
 \end{split}
\end{equation}
The $\alpha_{i,j}$ are coefficients relating stabilizer operators with specific projectors.
Lowercase Latin letters as well as $\sigma$, stand for one of the four single qubit Pauli operators $\{1, X, Y, Z\}$.
Thus $\sigma^L$, is a logical operator on the code and as a stabilizer code can be expressed as a tensor product of single qubit operators.
Finally $\overline{\Lambda}$ denotes the super-operator dual to $\Lambda$ (i.e. if $\Lambda(\rho) = \sum_{k} A_k \rho A_k^\dagger$, then $\overline{\Lambda}(O) = \sum_{k} A_k^\dagger O A_k$).

We may now give the analogous definitions for the case of an $M$ level concatenated code.
Here, objects must be further specified by a vector $\vec{v}$ of at most $M$ components indicating the physical qubit or group of qubits they act on.
Some of these objects require a base case definition for $\mid \vec{v} \mid = M$,
\begin{equation}
\begin{split}
\sigma^L_{\vec{v}} & =  \sigma_{\vec{v}} \\
 \sigma^{EC}_{\vec{v}} & =  \sigma_{\vec{v}} \\
 {\mathcal F}_{\vec{v}} &= {\mathbbm 1}.
\end{split}
\end{equation}
In this case, ${\vec{v}}$ identifies on which physical qubit(s) the operators act on.
The super-operator ${\mathcal F}$ represents the full recovery operation which is trivial in the case of physical qubits.
For the rest of the objects, definitions are only required for $\mid \vec{v} \mid < M$.
\begin{equation}
 \begin{split}
 S^{(j)}_{\vec{v}} & =  (s1^{(j)})_{1:\vec{v}}^{EC} \otimes \ldots \otimes (sk^{(j)})_{k:\vec{v}}^{EC} \\
 P^{(j)}_{\vec{v}} & =  \sum \alpha_{i,j} S^{(i)}_{\vec{v}} \\
 R^{(j)}_{\vec{v}} & =  (r1^{(j)})_{1:\vec{v}}^L \otimes \ldots \otimes (rk^{(j)})_{k:\vec{v}}^L \\
 {\mathcal R}_{\vec{v}}(\rho) & = \sum R^{(k)}_{\vec{v}}P^{(k)}_{\vec{v}} \rho P^{(k)}_{\vec{v}}R^{(k)}_{\vec{v}} \\
 \sigma^L_{\vec{v}} & =  (\sigma1)_{1:\vec{v}}^L \otimes  \ldots \otimes (\sigma k)_{1:\vec{v}}^L \\
 \sigma^{EC}_{\vec{v}} & =  \overline{\mathcal F}_{\vec{v}}(\sigma_{\vec{v}}^L)\\
 {\mathcal F}_{\vec{v}} &= {\mathcal R}_{\vec{v}} \circ \left( {\mathcal F}_{1:\vec{v}} \otimes \ldots \otimes {\mathcal F}_{k:\vec{v}}\right)
 .
\end{split}
\end{equation}
The main distinction from non-concatenated definitions is that the subindex $\vec{v}$ has been incorporated everywhere.
In addition, tensor product decomposition of operators now runs either in terms of logical operators (super-index $L$) or error corrected observables (super-index $EC$).
Finally, a distinction is made between ${\mathcal F}_{\vec{v}}$, which corrects all errors in a given block of qubits denoted by ${\vec{v}}$ and ${\mathcal R}_{\vec{v}}$  which corrects for only the highest level errors within that block.
This distinction may seem artificial since for a simple code (i.e. $\mid \vec{v} \mid = M-1$) a full correction corresponds to correcting the highest level error blocks possible.

We will now concentrate on some of the properties these recursive definitions carry that will later allow us to define the dissipative concatenated QECC and prove robustness results.
The main property relating logical and error corrected operators verified by definition is
\begin{equation}
 \tmop{tr}[\sigma^{EC}_{\vec{v}} \rho ] = \tmop{tr}[\sigma^L_{\vec{v}}{\mathcal F}_{\vec{v}}(\rho)].
\end{equation}
The meaningfulness of error corrected operators thus stems from the fact that if relatively few errors are applied to an encoded state, the error corrected operator provides the same expectation value as the logical operator on the unerred state
\begin{equation}
  \tmop{tr} [O^{E C}_{\vec{v}} \mathcal{E}(\rho)] = \tmop{tr} [O^L_{\vec{v}} \rho ] \hspace{1em} \forall \rho \in \tmop{codespace}_{\vec{v}},
\end{equation}
provided that the error super-operator $\mathcal{E}$ contains only ``few error'' Kraus operators.
More precisely, the expectation values are equal provided that
the Kraus operators for $\mathcal{E}$ contain less than $\lfloor \frac{d +
1}{2} \rfloor^{M - \mid \vec{v} \mid}$ errors.
More can be said in terms of
the structure of correctable errors.
Namely, there is a constant error threshold provided a random distribution of uncorrelated errors is assumed.

Another key property which can be guaranteed inductively is that
the commutation/anticommutation relation between logical operators and error
corrected observables should be the same as between bare operators.
\begin{equation}
  [\sigma1^{E C}_{\vec{v}}, \sigma2^L_{\vec{v}}]_{\pm} = 0 \hspace{1em} \Leftrightarrow
  \hspace{1em} [\sigma1, \sigma2]_{\pm} = 0 \label{CommutationECL}
\end{equation}
An even stronger statement can be made about products of logical operators (error corrected observables)
\begin{equation}
 \sigma1 \sigma2 = \theta \sigma3 \Rightarrow
 \left\{ \begin{array}{l}
  \sigma1^{L}_{\vec{v}} \sigma2^{L}_{\vec{v}} = \theta \sigma3^{L}_{\vec{v}} \text{ and } \\
  \sigma1^{EC}_{\vec{v}} \sigma2^{EC}_{\vec{v}} = \theta \sigma3^{EC}_{\vec{v}}
 \end{array} \right.,
\end{equation}
where $\theta$ is a phase in $\{1, -1, i, -i\}$.

The projector operators at each level are related to the presence of logical errors at the immediately preceding level.
This can be seen through the identity
\begin{equation}
  P^{(j)}_{\vec{v}} = R^{(j)}_{\vec{v}} P^{(0)}_{\vec{v}} R^{(j)}_{\vec{v}} \qquad R^{(0)}_{\vec{v}} = {\mathbbm 1},
\end{equation}
which relates $P^{(0)}_{\vec{v}}$, the trivial syndrome projector to other syndrome projectors.
The relation of this projector with the recovery operations is captured by
\begin{equation}
 P^{(0)}_{\vec{v}} \mathcal{R}_{\vec{v}} (\rho) P^{(0)}_{\vec{v}} =
   \mathcal{R}_{\vec{v}} (\rho).
\end{equation}

\emph{The master equation.-} considered for a dissipative protection on a concatenated QECC will contain error terms ${\mathcal D}_{noise, \vec{v}}$ on single physical qubits $\vec{v}$ as well as correction terms corresponding to each of the blocks.
The full master equation reads
\begin{equation}\label{eq:DissipativeDynamics}
 \dot{\rho} = \sum_{\mid \vec{v} \mid =M} {\mathcal D}_{noise, \vec{v}}(\rho) +
 \sum_{\mid \vec{v} \mid  < M} {\mathcal D}_{correct, \vec{v}}(\rho).
\end{equation}
Error terms ${\mathcal D}_{noise, \vec{v}}$ are single qubit superoperators with norm bounded by
$\left\| {\mathcal D}_{noise, \vec{v}} \right\| \leq \Gamma_{noise}$.
The protective dissipation ${\mathcal D}_{correct, \vec{v}}$ is defined by
\begin{equation}
 {\mathcal D}_{correct, \vec{v}}(\rho) = \Gamma_{correct, \vec{v}}[ {\mathcal R}_{\vec{v}}(\rho)-\rho ]
\end{equation}
which can be written in Lindblad form as
\begin{equation}
 {\mathcal D}_{correct, \vec{v}}(\rho) =
 \sum_{(j)} L^{(j)}_{\vec{v}} \rho L^{(j)\dagger}_{\vec{v}} -
 \frac{1}{2} \left\{ L^{(j)\dagger}_{\vec{v}} L^{(j)}_{\vec{v}}, \rho \right\}_+
\end{equation}
with Lindblad operators
\begin{equation}
L^{(j)}_{\vec{v}} = \sqrt{\Gamma_{correct, \vec{v}}} R^{(j)}_{\vec{v}}P^{(j)}_{\vec{v}}.
\end{equation}

We will prove the robustness of the highest level observables $X^{EC}_\emptyset, Y^{EC}_\emptyset, Z^{EC}_\emptyset$ under the combination of weak local noise and this dissipative protection.
To do this, we focus on the observables $\{P_{\vec{v}}^{(j)} : | \vec{v} | < M\}$.
Together with an arbitrary error corrected observable at the highest level, these constitute a complete set of quantum numbers.
The most attractive features of these observables is that both single qubit Pauli errors
and the recovery operations may be described by classical deterministic
transition rules in terms of this specific set of quantum numbers.
Furthermore, the events influencing each of these quantum numbers may be
simply characterized.
Namely, only recovery or physical error events located at $\vec{u} \succcurlyeq \vec{v}$ can influence the validity of $P_{\vec{v}}^{(j)}$.
This will allow us to provide upper bounds for the probability of introducing logical errors.

It is useful to define certain additional projectors in terms of the set of commuting projectors $\{P_{\vec{v}}^{(j)} : M > | \vec{v} |\}$
\begin{eqnarray}
  \tmop{HasError} ( \vec{v}) & = & \mathbbm{1}- P_{\vec{v}}^{(0)} \nonumber\\
  \tmop{IsError} (j : \vec{v}) & = & P_{\vec{v}}^{(X_{j})} +
  P_{\vec{v}}^{(Y_{j})} + P_{\vec{v}}^{(Z_{j})} \\
  \tmop{Enabled} (j : \vec{v}) & = & \mathbbm{1}- P_{\vec{v}}^{(0)} -
  P_{\vec{v}}^{(X_{j})} - P_{\vec{v}}^{(Y_{j})} - P_{\vec{v}}^{(Z_{j})} \nonumber
\end{eqnarray}
The recovery operation $\mathcal{R}_{\vec{v}}$ has a non
trivial effect only for the subspace ``$\tmop{HasError} ( \vec{v})$''.
Furthermore, in the subspace ``$\tmop{IsError} (j : \vec{v})$'', the effect of applying the recovery operation $\text{$\mathcal{R}_{\vec{v}}$}$ is to apply a logical operation on $j : \vec{v}$.
Finally, the projector ``$\tmop{Enabled} (j : \vec{v})$'' the difference between the two and indicates that there is already a logical error among the immediate components of $\vec{v}$, but that it is not at $j : \vec{v}$.
This last projector will be instrumental in bounding the probability for physical
errors to be raised as logical errors.
In the case of the perfect five qubit code, it is a necessary and sufficient condition for a logical operation at $j : \vec{v}$ be seen (in terms of the stabilizers) as raising a logical operation at $\vec{v}$.
The following short hand notation will be used to express the probability of satisfying these predicates (projectors)
\begin{equation}
  \langle P \rangle_t = \tmop{tr} [ P \rho (t)].
\end{equation}

\subsection{Bounding error probabilities}
of constitute the core of proving the robustness of error corrected observables under such a dissipative dynamics as Eq. (\ref{eq:DissipativeDynamics}).
In particular, we wish to prove inductively that
\begin{equation}
  \forall t \hspace{1em} \langle \tmop{HasError}( \vec{v}) \rangle_t \leq p_n
  \hspace{1em} \tmop{where} n = M - | \vec{v} |.
\end{equation}
Since the initial state $\rho (0)$ is by Hypothesis a code state at all
levels, we have that
\begin{equation}
  \forall \vec{v} : \hspace{1em} \langle \tmop{HasError} ( \vec{v}) \rangle_{t=0} = 0.
\end{equation}
The trick now is to obtain an upper bound on the rate at which these
probabilities may increase and upper-bound the actual probability by a
fixed-point value. Let us first illustrate this method by considering a simple
example provided by $| \vec{v} | = M - 1$.
\begin{equation}\begin{split}
  & \frac{d \langle \tmop{HasError} ( \vec{v}) \rangle_t }{d t} \leq \\
  & k \Gamma_{\tmop{noise}} - \Gamma_{\tmop{correct}, \vec{v}}
  \langle \tmop{HasError} ( \vec{v}) \rangle_t
\end{split}\end{equation}
Note that we have excluded processes by which a physical error cancels a
preexisting error.

From the rate bound, we may extract a fixed-point upper-bound and use it to
bound the actual probability
\begin{equation}
  \langle \tmop{HasError} ( \vec{v}) \rangle_t \leq \frac{k \Gamma_{\tmop{noise}}}{k \Gamma_{\tmop{noise}} + \Gamma_{\tmop{correct},
  \vec{v}}} .
\end{equation}
Assuming $\Gamma_{\tmop{correct}, \vec{v}} \geq \Gamma_{\tmop{correct}, M - | \vec{v} |}$, we may further simplify the bound to
\begin{equation}
  \langle \tmop{HasError} ( \vec{v}) \rangle_t \leq \frac{k
  \Gamma_{\tmop{noise}}}{\Gamma_{\tmop{correct}, 1}} =: p_1 .
\end{equation}
We may take a similar approach to bound the rate at which errors accumulate at
higher levels (i.e. $M - | \vec{v} | = n + 1$).
However, the expressions required here are a bit more complicated.
\begin{eqnarray}
  &  & \frac{d \langle \tmop{HasError} ( \vec{v}) \rangle_t }{d t}  \\
  & \leq & \sum_{{\begin{subarray}{c} \vec{u} \succ \vec{v}\\ | \vec{u} | = M \end{subarray}}} \Gamma_{\tmop{noise}, \vec{u}}
  \left\langle \prod_{\vec{u} \succcurlyeq \vec{w} \succ\succ \vec{v}}
  \tmop{Enabled} ( \vec{w})) \right\rangle_t  \label{exp:ProbForAll}\\
  &&- \Gamma_{\tmop{correct}, \vec{v}} \langle \tmop{HasError} ( \vec{v}) \rangle_t  \nonumber\\
  & \leq & \Gamma_{\tmop{noise}} \sum_{{\begin{subarray}{c} \vec{u} \succ \vec{v}\\| \vec{u} | = M \end{subarray}}}  \prod_{\vec{u} \succcurlyeq   \vec{w} \succ\succ \vec{v}}
  \langle \tmop{Enabled} ( \vec{w}) \rangle_t \label{exp:ProdProbs}\\
  & &- \Gamma_{\tmop{correct}, \vec{v}} \langle \tmop{HasError} ( \vec{v}) \rangle_t \nonumber\\
  & \leq & \Gamma_{\tmop{noise}}
  \sum_{{\begin{subarray}{c} \vec{u} \succ \vec{v}\\  | \vec{u} | = M \end{subarray}}}
  \prod_{\vec{u} \succ \vec{w} \succ \vec{v}}
  \langle \tmop{HasError} ( \vec{w}) \rangle_t \\
  & &- \Gamma_{\tmop{correct}, \vec{v}} \langle \tmop{HasError} ( \vec{v}) \rangle_t \nonumber\\
  & \leq & k^{n + 1} \Gamma_{\tmop{noise}} \prod^n_{j = 1} p_j \label{exp:final}\\
  && - \Gamma_{\tmop{correct}, n + 1} \langle \tmop{HasError} ( \vec{v}) \rangle \nonumber
\end{eqnarray}
A non trivial step is taken in going from [\ref{exp:ProbForAll}] to
[\ref{exp:ProdProbs}], where the probability of a conjunction is taken to be a
product of probabilities (i.e. independent probabilities).
This property will be proven in appendix \ref{app:IndependenceProof}.

In turn, this leads to the fixed point bound
\begin{equation}
  \langle \tmop{HasError} ( \vec{v}) \rangle_t \leq \frac{k^{n + 1}
  \Gamma_{\tmop{noise}} \prod^n_{j = 1} p_j}{\Gamma_{\tmop{correct}, n + 1}} =
  : p_{n + 1} .
\end{equation}
From here, we inductively derive the expression
\begin{equation}
  p_n = \frac{\Gamma_{\tmop{noise}}^{2^{n - 1}} k^{2^n - 1} }{
  \Gamma_{\tmop{correct}, n} \prod^{n - 1}_{j = 1} \Gamma_{\tmop{correct}, j}^{2^{n - 1 - j}}}
  .
\end{equation}
Making the additional assumption $\Gamma_{\tmop{correct}, j} =
\Gamma_{\tmop{correct}} \delta^j$ we may simplify this expression to obtain
\begin{equation}
  p_n = \left( \frac{\Gamma_{\tmop{noise}} k^2}{\Gamma_{\tmop{correct}}
  \delta^2} \right)^{2^{n - 1}} \frac{\delta}{k} .
\end{equation}
In turn, this tells us that if $\Gamma_{\tmop{noise}} < (\delta/k)^2\Gamma_{\tmop{correct}}$, then the probability of having non trivial syndrome decreases double exponentially with the level of the
syndrome.

Our final goal is to obtain an expression bounding the rate at which logical
errors are introduced. One possibility, is to study the decay rate for any of
the three highest level logical Pauli observables.
Since these three constitute a full set of observables for the logical subsystem, their preservation implies high fidelity storage of quantum information \cite{alicki_decay_2009}.

A logical error or flip of the highest level logical observables, can be introduced whenever a physical error occurs at a site which is enabled to raise the error at all levels.
Employing bounds similar to those in Eqs. (\ref{exp:ProbForAll})-(\ref{exp:final}) one arrives at
\begin{eqnarray}\label{eq:BoundECObservable}
  &  & \frac{d \langle X^{\tmop{EC}}_{\emptyset} \rho (t) \rangle_t }{d t} \\
  & \leq & \sum_{| \vec{u} | = M } \Gamma_{\tmop{noise}, \vec{u}}
  \left\langle \prod_{\vec{u}
  \succcurlyeq \vec{w} \succ \emptyset} \tmop{Enabled} ( \vec{w}) \right\rangle_t \\
  & \leq & \Gamma_{\tmop{noise}} \delta^M \left( \frac{\Gamma_{\tmop{noise}}
  k^2}{\Gamma_{\tmop{correct}} \delta^2} \right)^{2^M - 1},
\end{eqnarray}
indicating that for a sufficiently low physical error rate, the logical error rate is suppressed double exponentially in terms of $M$, similar to results for concatenated QECC in a quantum circuit model.

\section{Proof of independence for the Enabled property\label{app:IndependenceProof}}

In this section we assume a Pauli noise model and prove that the Enabled property along the different truncations of the same physical address are statistically independent.
More specifically, the factorization
\begin{equation}\label{eq:EnabledFactorization}
  \left\langle \prod_{\vec{u} \succcurlyeq \vec{w} \succ \vec{v}}
  \tmop{Enabled} ( \vec{w}) \right\rangle_t = \prod_{\vec{u} \succcurlyeq \vec{w}
  \succ \vec{v}} \langle \tmop{Enabled} ( \vec{w}) \rangle_t
\end{equation}
holds for a noise process of Pauli form
\begin{equation}\label{eq:PauliNoise}
{\mathcal D}_{noise, \vec{v}}(\rho) = \sum_{\sigma \in \{X,Y,Z\}} \Gamma_\sigma \left( \sigma_{\vec{v}} \rho \sigma_{\vec{v}} - \rho \right).
\end{equation}
The restriction of the noise process to Pauli form Eq. (\ref{eq:PauliNoise}) is clearly undesired.
However, it provides a sufficient condition to prove Eq. (\ref{eq:EnabledFactorization}), which does not hold for general noise.
We expect the need for this assumption to be an artifact of our proof technique and that our main result, i.e. Eq. (\ref{eq:BoundECObservable}), will essentially hold for any independent noise model.

The proof relies on the independence of the different processes which
introduce physical errors and perform recovery operations.
An event $\tmop{Ev}_{\vec{w}}$ will be associated to each vector $\vec{w}$,  with $| \vec{w} | = M$  corresponding to the introduction of physical errors at $\vec{w}$ and $| \vec{w} | < M$ corresponding to recovery operation $\mathcal{R}_{\vec{w}}$.
Each event $\tmop{Ev}_{\vec{w}}$ can be seen as the state dependent application of
a tensor product Pauli operator.
Furthermore, for $|\vec{w}|<M$ the operator only depends on the quantum numbers $P_{\vec{w}}^{(j)}$ and must be a logical Pauli operators at some $w_0:\vec{w}$.
The correction operators satisfy this property by design.
In turn, for $|\vec{w}|=M$, $\tmop{Ev}_{\vec{w}}$ applies a randomly chosen physical Pauli operator at $\vec{w}$ according to the Pauli form noise model Eq. (\ref{eq:PauliNoise}).
It can be seen that under these conditions, only events $\tmop{Ev}_{\vec{w}}$ such that $\vec{w} \succ \vec{v}$ can directly affect the quantum numbers $P_{\vec{v}}^{(j)}$.
Thus, given a history $L$ of events $\tmop{Ev}_{\vec{w}}$ applied to an initially encoded state,  the quantum numbers $P^{(j)}_{\vec{v}}$ are well defined and depend only on the sub-history of events $L'$ containing the events $\tmop{Ev}_{\vec{w}}$ with $\vec{w}\succcurlyeq \vec{v}$.

Since $\tmop{Enabled} (v_0 : \vec{v})$ can be defined in terms of the $P^{(j)}_{\vec{v}}$ it may only depend on the sub-history of events $\tmop{Ev}_{\vec{w}}$ with $\vec{w}\succcurlyeq \vec{c}$.
Furthermore, $\tmop{Enabled} (v_0 : \vec{v})$ will be shown not to depend direct or indirectly on  events $\tmop{Ev}_{\vec{u}}$ with $\vec{u} \succcurlyeq v_0 : \vec{v}$.
This can be seen as a consequence of $\tmop{Enabled} (v_0 : \vec{v})$ commuting with any Pauli operator acting on qubits $\vec{w}$ with $\vec{w}\succcurlyeq v_0 : \vec{v}$.

Proving Eq. (\ref{eq:EnabledFactorization}) may be split in the following steps
\begin{align}
 & \left\langle \prod_{\vec{u} \succcurlyeq \vec{w} \succ \vec{v}} \tmop{Enabled} ( \vec{w}) \right\rangle_t \\
  = &\sum_L p_L(t) \tmop{tr}[ \prod_{\vec{u} \succcurlyeq \vec{w} \succ \vec{v}} \tmop{Enabled} ( \vec{w}) L \rho_0] \label{exp:Unraveling}\\
  =& \sum_L p_L(t) \prod_{\vec{u} \succcurlyeq \vec{w} \succ \vec{v}} \tmop{tr}[  \tmop{Enabled} ( \vec{w}) L \rho_0] \label{exp:WellDefinedprojectors}\\
  =& \sum_L p_L(t) \prod_{\vec{u} \succcurlyeq \vec{w} \succ \vec{v}} \tmop{tr}[  \tmop{Enabled} ( \vec{w}) L_{\vec{w}} \rho_0] \label{exp:IndependentEvents}\\
    =& \prod_{\vec{u} \succcurlyeq \vec{w} \succ \vec{v}} \sum_{L_{\vec{w}}} p_{L_{\vec{w}}}(t) \tmop{tr}[  \tmop{Enabled} ( \vec{w}) L_{\vec{w}} \rho_0] \label{exp:IndependentEvents2}\\
  =& \prod_{\vec{u} \succcurlyeq \vec{w} \succ \vec{v}}
  \langle \tmop{Enabled} ( \vec{w}) \rangle_t ,\label{exp:product}
\end{align}
 which will be subsequently explained and justified.
As a first step, the master equation  defining $\rho(t)$ is unraveled \cite{carmichael_statistical_1998} into event histories $L$ to obtain Exp. (\ref{exp:Unraveling}).
Given that every event history $L$ implements a Pauli operator which produces eigenstates to all the projectors $\tmop{Enabled}(\vec{w})$, the $0, 1$ expectation values may be factorized to obtain Exp. (\ref{exp:WellDefinedprojectors}).
Expectation values depend only on disjoint sub-histories $L_{w_0:\vec{w}}$ ( a history of events uniquely determined by filtering events $\tmop{Ev}_{\vec{u}}$ such that $\vec{u}\succ \vec{w}$ but not $\vec{u}\succ w_0:\vec{w}$  from $L$ ), leading to Exp. (\ref{exp:IndependentEvents}).
Furthermore, the sum of $p_L$ consistent with given sub-histories $L_{\vec{w}}$ may be written as a product of the independent probabilities $p_{L_{\vec{w}}}$ of such sub-histories, thus leading to Exp. (\ref{exp:IndependentEvents2}).
Finally, each factors in Exp. (\ref{exp:IndependentEvents}) may be seen to be the history unraveling of each of the factors in Exp. (\ref{exp:product}), which is what we set out to prove.
\end{document}